\documentclass[aps,prl,twocolumn,superscriptaddress]{revtex4}
\usepackage{amsfonts}
\usepackage{amsmath}
\usepackage{graphicx}
\usepackage{dsfont}
\usepackage{diagbox}
\usepackage{array}
\usepackage{amssymb}
\usepackage{bbm}
\usepackage{float}
\usepackage{subfigure}
\usepackage{url}
\usepackage{hyperref}
\usepackage{graphicx}
\usepackage{epsfig}
\usepackage{color}

\newcommand{\PreserveBackslash}[1]{\let\temp=\\#1\let\\=\temp}
\newcolumntype{C}[1]{>{\PreserveBackslash\centering}p{#1}}
\newcolumntype{R}[1]{>{\PreserveBackslash\raggedleft}p{#1}}
\newcolumntype{L}[1]{>{\PreserveBackslash\raggedright}p{#1}}

\begin{document}

\title{Distinct pairing symmetries of superconductivity in infinite-layer
nickelates}
\author{Zhan Wang}
\affiliation{Kavli Institute for Theoretical Sciences and CAS Center for Topological
Quantum Computation, University of Chinese Academy of Sciences, Beijing
100190, China }
\author{Guang-Ming Zhang}
\email{gmzhang@tsinghua.edu.cn}
\affiliation{State Key Laboratory of Low-Dimensional Quantum Physics and Department of
Physics, Tsinghua University, Beijing 100084, China}
\affiliation{Frontier Science Center for Quantum Information, Beijing 100084, China}
\author{Yi-feng Yang}
\email{yifeng@iphy.ac.cn}
\affiliation{Beijing National Lab for Condensed Matter Physics and Institute of Physics,
Chinese Academy of Sciences, Beijing 100190, China}
\affiliation{School of Physical Sciences, University of Chinese Academy of Sciences,
Beijing 100190, China}
\affiliation{Songshan Lake Materials Laboratory, Dongguan, Guangdong 523808, China}
\author{Fu-Chun Zhang}
\email{fuchun@ucas.ac.cn}
\affiliation{Kavli Institute for Theoretical Sciences and CAS Center for Topological
Quantum Computation, University of Chinese Academy of Sciences, Beijing
100190, China }
\date{\today }

\begin{abstract}
We report theoretical predictions on the pairing symmetry of the newly
discovered superconducting nickelate Nd$_{1-x}$Sr$_{x}$NiO$_{2}$ based on
the renormalized mean-field theory for a generalized model Hamiltonian
proposed in [Phys. Rev. B \textbf{101}, 020501(R)]. For practical values of
the key parameters, we find a transition between a gapped ($d+is$)-wave
pairing state in the small doping region to a gapless $d$-wave pairing state
in the large doping region, accompanied by an abrupt Fermi surface change at
the critical doping. Our overall phase diagram also shows the possibility of
a ($d+is$)- to $s$-wave transition if the electron hybridization is
relatively small. In either case, the low-doping ($d+is$)-wave state is a
gapped superconducting state with broken time-reversal symmetry. Our results
are in qualitative agreement with recent experimental observations and
predict several key features to be examined in future measurements.
\end{abstract}

\maketitle

\textbf{Introduction.} -The recent discovery of superconductivity (SC) in
the single crystal thin films of infinite-layer nickelates Nd$_{1-x}$Sr$_{x}$%
NiO$_{2}$ \cite{Li2019} has stimulated intensive debates on its underlying
electronic structural properties and superconducting pairing symmetries \cite%
{Norman2019,Sawatzky2019,Kuroki2019,Hepting2019,Normura2019,Thomale,zhang-yang-zhang,Hanghui2020,Millis2020}%
. Despite the similarities in the crystal structure and $3d^{9}$
configuration of the nickelate and cuprate superconductors, there are
increasing evidences suggesting that these two systems might belong to
different classes of unconventional SC. Earlier first-principles
calculations have revealed subtle differences in their band structures \cite%
{Anisimov1999,Hayward1999,Lee2004,Botana2017,Chaloupka2008,Hansmann2009,Middey2016,Boris2011,Benckiser2011,Disa2015,Zhang2017}%
. In experiments, the parent compound NdNiO$_{2}$ displays paramagnetic
metallic behavior at high temperatures with a resistivity upturn below about
$70$ K, showing no sign of any magnetic long-range order \cite{Hayward2003}.
This is in stark contrast with the cuprates whose parent compound is a
charge-transfer insulator with antiferromagnetic (AF) long-range order. As a
consequence, the nickelates may be modelled as a self-doped Mott insulator
with two types of charge carriers \cite{zhang-yang-zhang}, with the
low-temperature upturn \cite{Li2019,Ikeda2016} arising from the Kondo
coupling between low-density conduction electrons and localized Ni-3$%
d_{x^{2}-y^{2}}$ moments \cite{Foot-Note}. This produces both Kondo singlets
(doublons) and holes moving through the lattice of otherwise nickel spin-1/2
background, suppressing the AF long-range order, and causing a phase
transition to a paramagnetic metal \cite{zhang-yang-zhang}. Latest
measurements \cite{Hepting2019,Kourkoutis} and first-principle calculations%
\cite{Lee2004,Hanghui2020} confirm this scenario and reveal a special
interstitial $s$ orbital for the hybridization \cite{Hanghui2020}, which is
missing in previous calculations.

We expect these differences to have an immediate impact on the candidate
pairing mechanism. In cuprates, additional holes are doped on the oxygen
sites in the CuO$_{2}$ planes \cite{Muller,Anderson,AtoZ,LeeNagaosaWen} and
combine with the $3d_{x^{2}-y^{2}}$ spins of Cu-ions to form the Zhang-Rice
singlets \cite{ZhangRice}. High temperature SC with robust $d$-wave pairing
can be derived from an effective one-band $t$-$J$ model \cite%
{Shen,Harlingen,Tsuei}. In nickelates, Sr doping may not only introduce
additional holes on the oxygen sites to form the Ni-O spin singlets or
holons (a spin zero state) \cite{Sawatzky2019,WeiKu}, but also reduce the
number of conduction electrons, thus tilting the balance between the
electron and hole carriers of distinct characters. The Hall coefficient is
then expected to vary gradually and change sign with doping or temperature.
One thus anticipates more rich physics in the nickelate superconductors,
whose pairing symmetry may be altered by the hybridization. Indeed, latest
experiment has revealed a non-monotonic change of $T_{c}$ in exact
accordance with the sign change of the Hall coefficient \cite%
{Li2020,Zeng2020}.

To further elucidate the pairing symmetry of the nickelate superconductors,
we employ here the renormalized mean-field theory (RMFT) \cite{RMFT} and
study the superconducting pairing symmetry based on a generalized $K$-$t$-$J$
model \cite{zhang-yang-zhang} in Eq. (1) and Eq. (2). Our calculations lead
to a global phase diagram depending on the hole concentration $p$ and the
conduction electron hopping ($t_c/K$) which controls the effective strength
of the Kondo hybridization. At small doping and with reasonable choices of
parameters, our calculations reveal an unusual gapped ($d+is$)-wave SC with
the time-reversal symmetry breaking, which is distinctly different from the
familiar cuprate superconductivity. For large doping, we find either
extended $s$-wave pairing or pure $d$-wave pairing. The latter is quite
robust and occupies a large region in the phase diagram. Comparison with
experiment tends to favor a transition from the gapped ($d+is$)-wave to
gapless $d$-wave pairing states with increasing hole doping. We further
predict that the SC transition is accompanied with an abrupt Fermi surface
change associated with the breakdown of the Kondo hybridization, causing
potentially a crossover line in the temperature-doping phase diagram as
observed in recent Hall measurements \cite{Li2020,Zeng2020}.

\textbf{Model Hamiltonian and RMFT.} - We start by first introducing the
generalized $K$-$t$-$J$ model for the nickelate superconductors, given by $%
H=H_{t-J}+H_{K}$. The $t$-$J$ part describes the hole doped lattice of Ni $%
3d_{x^{2}-y^{2}}$ spins with the nearest-neighbor AF superexchange
interactions,
\begin{equation}
H_{t-J}=-\sum_{ij\sigma }\left( t_{ij}P_{G}d_{i\sigma }^{\dagger }d_{j\sigma
}P_{G}+\text{h\text{.c.}}\right) +J\sum_{\langle ij\rangle }S_{i}\cdot S_{j},
\end{equation}
where $d_{i\sigma }$ and $d_{i\sigma}^{\dag }$ are the annihilation and
creation operators of the Ni $3d_{x^{2}-y^{2}}$ electrons, respectively, $%
t_{ij}$ is the hopping integral between site $i$ and $j$, and $P_{G}$ is the
Gutzwiller operator to project out doubly occupied electron states on the Ni
sites. For simplicity, we consider only the nearest neighbor hopping (NN) $t$
and next-nearest neighbor (NNN) hopping $t^{\prime}$. The AF superexchange $%
J $ is induced by the O-$2p$ orbitals but greatly reduced compared to that
in cuprates. The Kondo hybridization part is given by,
\begin{equation}
H_{K}=-t_{c}\sum_{\langle ij\rangle ,\sigma }\left( c_{i\sigma }^{\dagger
}c_{j\sigma }+h.c.\right) +\frac{K}{2}\sum_{j\alpha ;\sigma \sigma ^{\prime
}}S_{j}^{\alpha }c_{j\sigma }^{\dagger }\tau _{\sigma \sigma ^{\prime
}}^{\alpha }c_{j\sigma ^{\prime }},
\end{equation}
where $c_{i\sigma }$ ($c_{i\sigma}^{\dag }$) are the annihilation (creation)
operators of the conduction electrons from Nd $5d$, interstitial $s$, or
other extended orbitals, $t_{c}$ describes the effective hoping amplitude of
the conduction electrons projected on the square lattice sites of the Ni$%
^{1+}$ ions, $\tau ^{\alpha }$ ($\alpha =x,y,z$) are the spin-1/2 Pauli
matrices, and $K$ is the effective Kondo exchange coupling.

In the parent compounds LnNiO$_{2}$ (Ln=Nd, La, Pr), the total electron
density ($n_{c}+n_{d}$) is one per unit cell, hence the total holon density $%
n_{h}=n_{c}$. For Sr doped compounds, the hole doping $p=n_{h}-n_{c}>0$.
Analyses of the Hall coefficients at high temperatures suggest that the
average number of the conduction electrons is always small, i.e., $%
n_{c}=N^{-1}\sum_{j\sigma }\langle c_{j\sigma }^{\dagger }c_{j\sigma
}\rangle \ll 1$, where $N$ is the total number of the lattice sites.

For the RMFT calculations, the Gutzwiller renormalization factor should be
included to approximate the projection operator that projects out the doubly
occupied states. We have $g_{t}=n_{h}/(1+n_{h})$ for the constraint electron
hopping $t$ and $t^{\prime }$, $g_{J}=4/(1+n_{h})^{2}$ for the the AF
Heisenberg exchange $J$, and $g_{K}=2/(1+n_{h})$ for the Kondo exchange
coupling $K$. Four different mean-field order parameters are then introduced
to decouple the quartic AF Heisenberg spin exchange and the Kondo exchange
interactions:
\begin{eqnarray}
\chi _{ij} &=&\langle d_{i\uparrow }^{\dagger }d_{j_{\uparrow
}}+d_{i\downarrow }^{\dagger }d_{j\downarrow }\rangle ,\quad B=\frac{1}{%
\sqrt{2}}\langle d_{j\uparrow }^{\dagger }c_{j\downarrow }^{\dagger
}-d_{j\downarrow }^{\dagger }c_{j\uparrow }^{\dagger }\rangle ,  \notag \\
\Delta _{ij} &=&\langle d_{i\uparrow }^{\dagger }d_{j_{\downarrow
}}^{\dagger }-d_{i\downarrow }^{\dagger }d_{j\uparrow }^{\dagger }\rangle
,\quad D=\frac{1}{\sqrt{2}}\langle c_{j\uparrow }^{\dagger }d_{j\uparrow
}+c_{j\downarrow }^{\dagger }d_{j\downarrow }\rangle .  \notag
\end{eqnarray}%
The resulting mean-field Hamiltonian has a bilinear form and can be
expressed in the momentum space,
\begin{equation}
\mathcal{H}_{\text{mf}}=\sum_{\mathbf{k}}\Psi _{\mathbf{k}}^{\dag }\left(
\begin{array}{cccc}
\chi (\mathbf{k}) & K_{D} & \Delta ^{\ast }(\mathbf{k}) & K_{B}^{\ast } \\
K_{D}^{\ast } & \epsilon _{c}(\mathbf{k}) & K_{B}^{\ast } & 0 \\
\Delta (-\mathbf{k}) & K_{B} & -\chi (-\mathbf{k}) & -K_{D}^{\ast } \\
K_{B} & 0 & -K_{D} & -\epsilon _{c}(-\mathbf{k})%
\end{array}%
\right) \Psi _{\mathbf{k}},  \notag
\end{equation}%
where the Nambu spinors are defined as $\Psi _{\mathbf{k}}^{\dagger }=(d_{%
\mathbf{k}\uparrow }^{\dagger },c_{\mathbf{k}\uparrow }^{\dagger },d_{-%
\mathbf{k\downarrow }},c_{-\mathbf{k}\downarrow })$, and the matrix elements
are%
\begin{eqnarray}
\chi (\mathbf{k}) &=&-\sum_{\alpha }\left( tg_{t}+\frac{3}{8}Jg_{J}\chi
_{\alpha }\right) \cos (\mathbf{k}\cdot \alpha )  \notag \\
&&-t^{\prime }g_{t}\sum_{\delta }\cos (\mathbf{k}\cdot \delta )+\mu _{1},
\notag \\
\epsilon _{c}(\mathbf{k}) &=&-t_{c}\sum_{\alpha }\cos (\mathbf{k}\cdot
\alpha )+\mu _{2},  \notag \\
\Delta (\mathbf{k}) &=&-\frac{3}{8}Jg_{J}\sum_{\alpha }\Delta _{\alpha }\cos
(\mathbf{k}\cdot \alpha ),  \notag \\
K_{D} &=&-\frac{3}{4}g_{K}K\frac{D}{\sqrt{2}},K_{B}=-\frac{3}{4}g_{K}K\frac{B%
}{\sqrt{2}}.
\end{eqnarray}%
Here $\alpha $ denotes the vectors of the NN lattice sites and $\delta $
stands for those of the NNN sites. $\mu _{1}$ and $\mu _{2}$ are the
chemical potentials fixing the numbers of the constraint electrons $%
d_{i\sigma }$ and conduction electrons $c_{i\sigma }$, respectively.

The above mean-field Hamiltonian can be diagonalized using the Bogoliubov
transformation, $\left( d_{\mathbf{k}\uparrow },c_{\mathbf{k}\uparrow },d_{-%
\mathbf{k}\downarrow }^{\dagger },c_{-\mathbf{k}\downarrow }^{\dagger
}\right) ^{T}=U_{\mathbf{k}}\left( \alpha _{\mathbf{k}\uparrow },\beta _{%
\mathbf{k}\uparrow },\alpha _{-\mathbf{k}\downarrow }^{\dagger },\beta _{-%
\mathbf{k}\downarrow }^{\dagger }\right) ^{T}$. The ground state is given by
the vacuum of the Bogoliubov quasiparticles $\{\alpha _{\mathbf{k}\sigma
}^{\dagger },\beta _{\mathbf{k}\sigma }^{\dagger }\}$, which in turn yields
the self-consistent mean-field equations:%
\begin{eqnarray}
B &=&\frac{1}{\sqrt{2}N}\sum_{\mathbf{k}}(u_{13}^{\ast \mathbf{k}}u_{43}^{%
\mathbf{k}}+u_{14}^{\ast \mathbf{k}}u_{44}^{\mathbf{k}}-u_{21}^{\ast \mathbf{%
k}}u_{31}^{\mathbf{k}}-u_{22}^{\ast \mathbf{k}}u_{32}^{\mathbf{k}}),  \notag
\\
D &=&\frac{1}{\sqrt{2}N}\sum_{\mathbf{k}}(u_{23}^{\ast \mathbf{k}}u_{13}^{%
\mathbf{k}}+u_{24}^{\ast \mathbf{k}}u_{14}^{\mathbf{k}}+u_{31}^{\ast \mathbf{%
k}}u_{41}^{\mathbf{k}}+u_{32}^{\ast \mathbf{k}}u_{42}^{\mathbf{k}}),  \notag
\\
\chi _{\alpha } &=&\frac{2}{N}\sum_{\mathbf{k}}\exp [i\mathbf{k}\cdot \alpha
](u_{13}^{\ast \mathbf{k}}u_{13}^{\mathbf{k}}+u_{14}^{\ast \mathbf{k}%
}u_{14}^{\mathbf{k}}),  \notag \\
\Delta _{\alpha } &=&\frac{2}{N}\sum_{\mathbf{k}}\exp [i\mathbf{k}\cdot
\alpha ](u_{13}^{\ast \mathbf{k}}u_{33}^{\mathbf{k}}+u_{14}^{\ast \mathbf{k}%
}u_{34}^{\mathbf{k}}),  \notag \\
n_{c} &=&\frac{1}{N}\sum_{\mathbf{k}}(u_{23}^{\mathbf{k}}u_{23}^{\ast
\mathbf{k}}+u_{24}^{\mathbf{k}}u_{24}^{\ast \mathbf{k}}+u_{41}^{\mathbf{k}%
}u_{41}^{\ast \mathbf{k}}+u_{42}^{\mathbf{k}}u_{42}^{\ast \mathbf{k}}),
\notag \\
1-n_{h} &=&\frac{1}{N}\sum_{\mathbf{k}}(u_{13}^{\mathbf{k}}u_{13}^{\ast
\mathbf{k}}+u_{14}^{\mathbf{k}}u_{14}^{\ast \mathbf{k}}+u_{31}^{\mathbf{k}%
}u_{31}^{\ast \mathbf{k}}+u_{32}^{\mathbf{k}}u_{32}^{\ast \mathbf{k}}),
\end{eqnarray}%
where $u_{ij}^{\mathbf{k}}$ are given by the matrix elements of $U_{\mathbf{k%
}}$, and the last two equations fix the chemical potentials $\mu _{1}$ and $%
\mu _{2}$, respectively.

\textbf{Numerical results.} -For clarity, we define $\Delta
_{s}=|\Delta_{x}+\Delta _{y}|/2$ and $\Delta _{d}=|\Delta _{x}-\Delta
_{y}|/2 $ to represent the $s$ and $d$-wave pairing amplitudes, respectively. To numerically solve these self-consistent equations, we first fix the practical parameters based roughly on the experimental analyses and first-principle results. The Kondo coupling $K$ is considered to be the largest energy scale and thus chosen as the energy unit ($K=1$). To simplify the discussions, only the numerical results for the NN hopping $t=0.2$, the NNN hopping $t^{\prime }=-0.05$, and the AF Heisenberg spin exchange $J=0.1$
are presented. The density of the conduction electrons is set to $n_{c}=0.1$. These parameters may vary among different systems but the qualitative physical features will not be changed.

First of all, the overall phase diagram is displayed in Fig. \ref%
{Phasediagram} with the values of $t_{c}/K$ and the hole concentration $p$.
We find a dominant $d$-wave pairing symmetry in the phase diagram, which,
for small $t_{c}/K$ and large doping, turns into an extended $s$-wave state.
Most intriguingly, we find a large region of the $\left( d+is\right) $-wave
pairing for small hole doping. This exotic pairing state breaks the
time-reversal symmetry and its presence reflects a unique feature of the
nickelate superconductivity due to the interplay of the Kondo and Mott
physics in comparison with the cuprates.
\begin{figure}[tbp]
\centering
\includegraphics[width=\columnwidth]{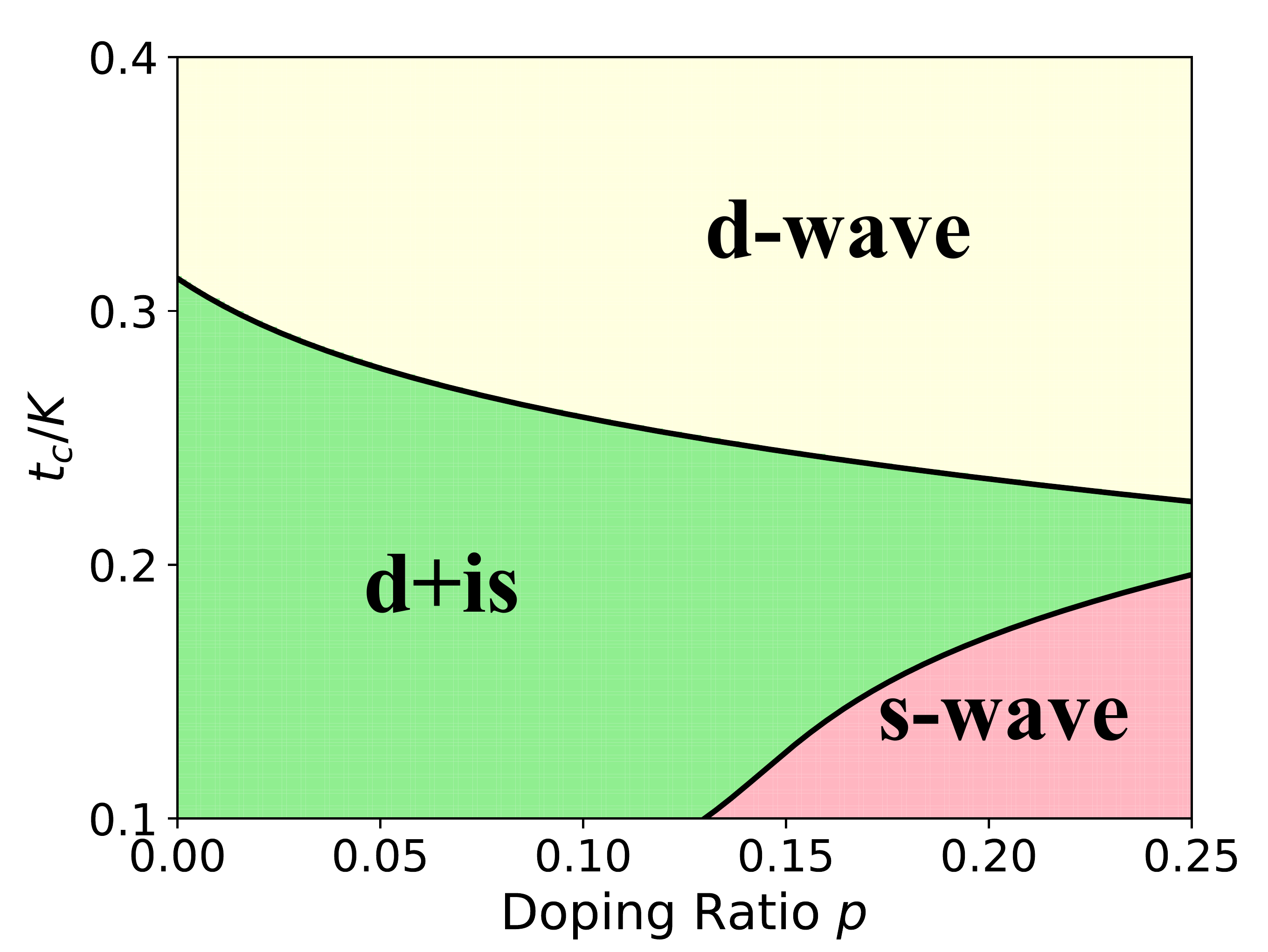}
\caption{Theoretical phase diagram of the superconductivity with varying
hopping $t_{c}/K$ and hole concentration $p$. At small doping, the pairing
symmetry is primarily ($d+is$)-wave SC. At large doping, the pairing
is either $s$-wave SC for small $t_{c}/K$ or $d$-wave SC for
large $t_{c}/K$.}
\label{Phasediagram}
\end{figure}

\begin{figure}[tbp]
\centering
\includegraphics[width=\columnwidth]{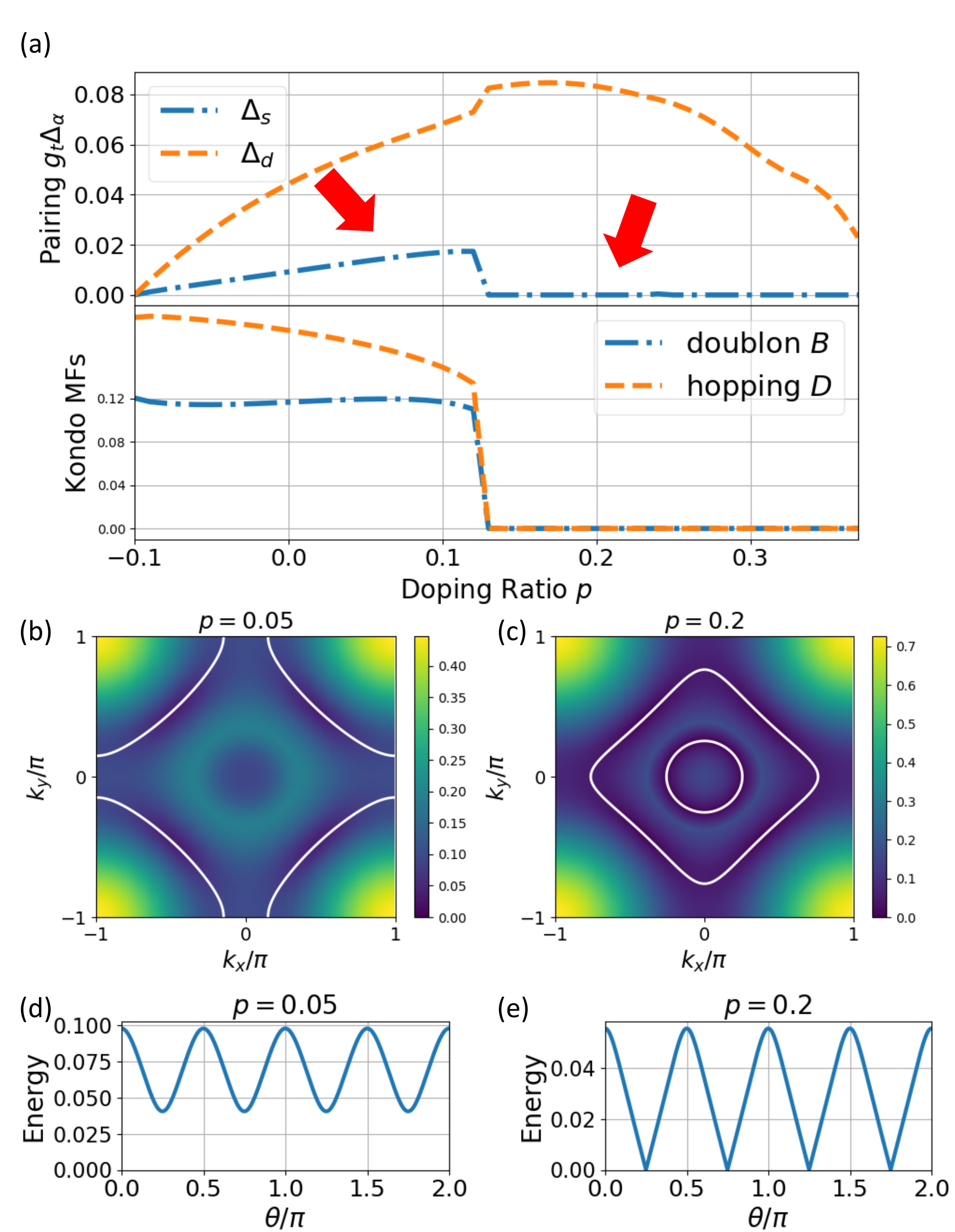}
\caption{RMFT results for $t_{c}/K=0.25$. (a) The mean-field parameters as a
function of doping for $g_{t}\Delta $ (upper panel) and $B$ and $D$ (lower
panel). (b) and (c) show the quasiparticle excitation energy (background)
and the Fermi surface (white solid line) defined as the minimal excitation
energy at $p=0.05$ ($d+is$)-wave and $p=0.2$ ($d$-wave) as marked by the
arrows in (a). (d) and (e) show the respective quasiparticle excitation gap
along the Fermi surface.}
\label{MF1}
\end{figure}

Details on the transition from the mixed $\left( d+is\right) $-wave SC to
the pure $d$-wave SC can be found in Fig. \ref{MF1}(a) for an intermediate $%
t_{c}/K=0.25$. The critical hole doping is $p^{\ast }\approx 0.13$, which is
comparable with the experiment but may vary with $t_{c}$ and other
controlling parameters. The transition is accompanied with vanishing Kondo
mean-field parameters $B$ and $D$, implying a breakdown of the Kondo
hybridization in the large doping side. It also implies that the $s$-wave
component is primarily associated with the Kondo hybridization effect and
the $d$-wave component is from the usual $t$-$J$ model. The corresponding
Fermi surface structures in these two different doping regions can be
extracted from the minimal energy contour of the SC quasiparticle excitation
energy. Two typical dopings for $p=0.05$ and $p=0.2$ are plotted in Figs. %
\ref{MF1}(b) and \ref{MF1}(c), and the corresponding SC gap functions are
plotted in Figs. \ref{MF1}(d) and \ref{MF1}(e). For small doping $p<p^{\ast
} $, the normal state has a large hole-like Fermi surface around four
Brillouin zone corners, while for large doping, two types of charge carriers
are effectively decoupled and give rise to two separate electron-like Fermi
surfaces around the Brillouin zone center. The physics of the pure $d$-wave
pairing region is similar to that of heavily hole-doped cuprates for this
particular doping. We have thus a concurrent Lifshitz transition at the
critical hole doping $p^{\ast }$, accompanying the transition between
different pairing states of the nickelate superconductivity.

\begin{figure}[tbp]
\centering
\includegraphics[width=\columnwidth]{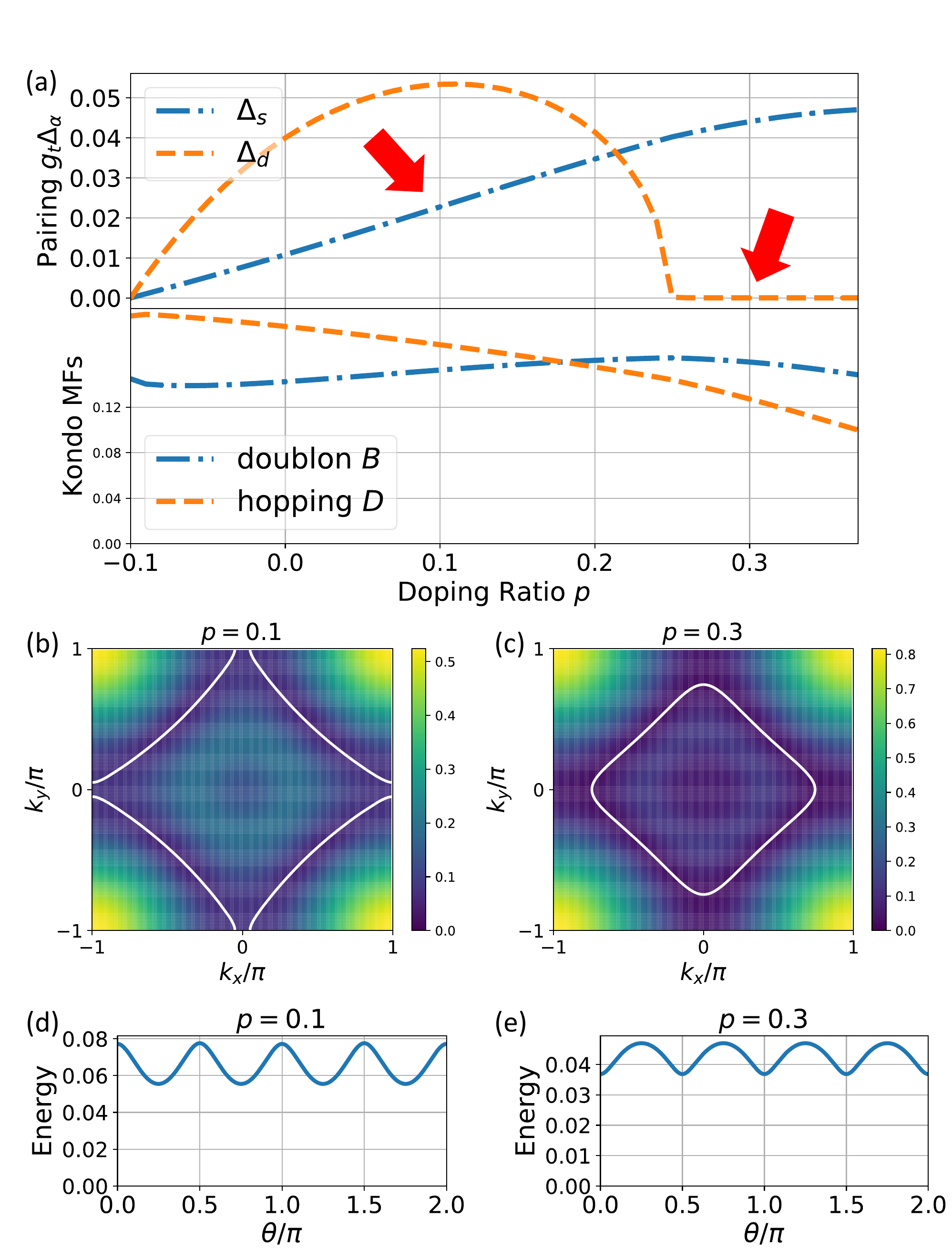}
\caption{RMFT results for $t_{c}/K=0.2$. (a) The mean-field parameters as a
function of doping for $g_{t}\Delta $ (upper panel) and $B$ and $D$ (lower
panel). (b) and (c) show the quasiparticle excitation energy (background)
and the Fermi surface (white solid line) at $p=0.1$ ($d+is$)-wave and $p=0.3$
($s$-wave) as marked by the arrows in (a). (d) and (e) show the respective
quasiparticle excitation gap along the Fermi surface.}
\label{MF2}
\end{figure}

For comparison, the RMFT results for a smaller $t_{c}/K=0.2$ are also
presented in Fig. \ref{MF2}, where the hole doping induces a transition from
the $\left( d+is\right) $-wave to the pure $s$-wave SC. In both phases, the
SC are gapped and the Kondo mean-field parameters $D$ and $B$ remain finite.
Hence the hybridization effect is not affected across the transition. Again,
the Fermi surfaces for $p=0.1$ and $p=0.3$ are extracted and shown in Figs. %
\ref{MF2}(b) and \ref{MF2}(c), respectively, with the corresponding SC gap
functions displayed in Figs. \ref{MF2}(d) and \ref{MF2}(e). We still see a
Lifshitz transition of the Fermi surfaces, but it is no longer associated
with the breakdown of the hybridization but a pure doping effect as in
cuprates.

\textbf{Discussions and Conclusion.} -The generalized $K$-$t$-$J$ model
contains several key energy scales that need to be fixed for better
experimental comparison in each individual compound. While the conduction
electron hopping $t_{c}$ may be roughly estimated from band calculations,
the constraint electron hoppings $t$ and $t^{\prime }$ are strongly
renormalized due to the background AF correlations. Following our previous
analyses, the Heisenberg superexchange $J$ is expected to be roughly the
order of 10-100 meV, which is smaller than that of cuprates due to the
larger charge transfer energy between O-$2p$ and Ni-$3d_{x^{2}-y^{2}}$
orbitals. The Kondo exchange interaction $K$ is estimated to be the order of
100-1000 meV \cite{zhang-yang-zhang, Hanghui2020}. This justifies our choice of $J/K$ in current numerical calculations. In any case, our results may serve as a qualitative guide for future studies on nickelate superconductors.

It is worthwhile comparing our results with the available experiment. Recent
systematic measurements on the resistivity and Hall coefficients in Nd$%
_{1-x} $Sr$_{x}$NiO$_{2}$ have revealed a non-monotonic doping dependence of
the superconducting $T_{c}$, whose local minimum coincides with the sign
change of the Hall coefficient \cite{Li2020,Zeng2020}. The latter further
gives rise to a crossover line in the temperature-doping phase diagram of
the nickelate superconductors. A straightforward comparison suggests that
the experimental observation may correspond to our derived transition from
the ($d+is$)-wave pairing to the $d$-wave or $s$-wave paring. The concurrent
change in the Hall coefficient therefore marks a potential Fermi surface
change, in resemblance of that observed in some heavy fermion systems owing
to the breakdown of the Kondo hybridization \cite{Si2014}. The latter also
leads to a delocalization line in the temperature-pressure or
temperature-doping phase diagram \cite{Yang2017}. It is thus attempted to
link the experiment with our theoretical proposals, predicting the SC
transition from a gapped ($d+is$)-wave state to a gapless $d$-wave pairing
state, with the crossover line in the temperature-doping plane potentially
associated with the Fermi surface change due to the Kondo hybridization.

If this is the case, one may further expect several key features to be
examined in future experiment: 1) a superconducting transition between
gapped and gapless pairings with increasing doping to be best revealed by
the scanning tunneling spectroscopy or the penetration depth measurement; 2)
time-reversal symmetry breaking in the low-doping gapped SC phase to be
detected in the $\mu $SR or Kerr experiments; 3) Fermi surface
reconstruction accompanying the superconducting transition to be measured by
the quantum oscillation experiments or angle-resolved photoemission
spectroscopy. Additionally, there may also exist other exotic properties
associated with the quantum critical point, besides the non-Fermi liquid
behavior which has been observed in superconducting nickelate thin films
with $\rho \sim T^{\alpha }$ and $\alpha =1.1-1.3$ \cite{zhang-yang-zhang}.
It will be interesting to see if future measurements will confirm these
preliminary predictions or suggest a different parameter region in our
generalized model.

In conclusion, we have discussed the pairing symmetry of the newly
discovered superconducting nickelate Nd$_{1-x}$Sr$_{x}$NiO$_{2}$ based on
the RMFT for a generalized $K$-$t$-$J$ model. Our calculations reveal an
interesting interplay between the Kondo and Mott physics. For practical
choices of the parameters, we find a transition from a gapped ($d+is$)-wave
state to a gapless $d$-wave state with increasing doping. An extended $s$%
-wave pairing has also been predicted but requires sufficiently small
hopping and large doping. For the former transition, our calculations
suggest a concurrent Fermi surface change and a corresponding crossover line
in the temperature-doping phase diagram due to the breakdown of the Kondo
hybridization. Our proposal is in good agreement with available experiments
and gives several key predictions for further verification.

\textbf{Note Added}. As we are finishing this manuscript, single particle
tunneling measurements \cite{HaihuWen} were reported on superconducting
nickelate thin films with $T_c\approx 9.1 K$, and two distinct types of
tunneling spectra were revealed: a V-shape feature with a gap maximum 3.9
meV, a U-shape feature with a gap about 2.35 meV, and some spectra with
mixed contributions of the two components. These spectra were ascribed to
different Fermi surfaces from the conduction and Ni $3d_{x^{2}-y^{2}}$
orbitals. However, according to our present calculations, these distinct
tunneling spectra observed at different locations on the thin films may be
caused by different hole doping concentrations due to surface effects, so
the different spectral shapes may correspond to the different pairing states
in our theory. In this sense, the tunneling experiment is supportive of our
theoretical prediction of multiple superconducting phases.

\textit{Acknowledgment}.- This work was supported by the National Key
Research and Development Program of MOST of China (2016YFYA0300300,
2017YFA0302902, 2017YFA0303103), the National Natural Science Foundation of
China (11774401, 11674278), the State Key Development Program for Basic
Research of China (2014CB921203 and 2015CB921303), and the Strategic
Priority Research Program of CAS (Grand No. XDB28000000).


\begin{thebibliography}{99}
\bibitem{Foot-Note} Actually an alternative explanation for the resistivity
upturn at low temperatures is weak localization due to the presence of
disorder holes in the NiO$_{2}$ plane. It can also give rise to a
logarithmic temperature dependent correction. However, the corresponding
correction to the Hall coefficient is independent of temperature in the same
region, which does not support this explanation.
\end{thebibliography}
\end{document}